\documentclass[conference]{IEEEtran}

\usepackage{graphicx} 
\DeclareGraphicsExtensions{.eps}
\usepackage{afterpage}
\usepackage{pgfplots} 
\usepackage{pgfplots,pgfplotstable}
\usepackage{graphicx}
\usepackage{subfigure} 
\usepackage{multicol} 
\usepackage{textgreek}
\usepackage{upgreek}
\usepackage{caption}
\pgfplotsset{compat=1.5}
\usepackage{subcaption} 
\usepackage[inkscapelatex=false]{svg}
\usepackage{multirow}
\usepackage{amsmath}
\usepackage{textcomp}
\usetikzlibrary{shapes,arrows}

\usepackage{empheq} 

\usepackage{stfloats} 

\usepackage[hyphens]{url}
\usepackage[hidelinks]{hyperref}
\hypersetup{colorlinks=true,linkcolor=black,citecolor=black,filecolor=black,urlcolor=black,breaklinks=true}
\usepackage{cite} 

\usepackage{color}

\usepackage{cancel} 

\usepackage{array} 

\usepackage[export]{adjustbox} 

\usepackage[normalem]{ulem} 

\usepackage{amsfonts}

\makeatletter
\def\bstctlcite{\@ifnextchar[{\@bstctlcite}{\@bstctlcite[@auxout]}}
\def\@bstctlcite[#1]#2{\@bsphack
 \@for\@citeb:=#2\do{%
   \edef\@citeb{\expandafter\@firstofone\@citeb}%
   \if@filesw\immediate\write\csname #1\endcsname{\string\citation{\@citeb}}\fi}%
 \@esphack}
\makeatother

\begin{document}

\bstctlcite{IEEEexample:BSTcontrol}

\title{Analyzing the Role of the DSO in Electricity Trading of VPPs via a Stackelberg Game Model}

\author{
\IEEEauthorblockN{Peng Wang\IEEEauthorrefmark{1}, Xi Zhang\IEEEauthorrefmark{2} and Luis Badesa\IEEEauthorrefmark{1}}
\vspace{3mm}
\IEEEauthorblockA{\IEEEauthorrefmark{1}\textit{School of Industrial Engineering and Design (ETSIDI), Technical University of Madrid (UPM), Madrid, Spain}
\\peng.wang@alumnos.upm.es, luis.badesa@upm.es}
\vspace{1mm}
\IEEEauthorblockA{\IEEEauthorrefmark{2}\textit{Department of Information and Communication, China Electric Power Research Institute, Beijing, China}
\\{x.zhang14@imperial.ac.uk}}
}

\maketitle

\begin{abstract}
The increasing penetration of distributed energy resources has sparked interests in participating in power markets. Here, we consider two settings where Virtual Power Plants (VPPs) with some flexible resources participate in the electricity trading, either directly in the wholesale electricity market, or interfaced by the Distribution System Operator (DSO) who is the transaction organizer. In order to study the role of DSO as a stakeholder, a Stackelberg game is represented via a bi-level model: the DSO maximizes profits at the upper level, while the VPPs minimize operating costs at the lower level. To solve this problem, the Karush-Kuhn-Tucker conditions of lower level is deduced to achieve a single-level problem. The results show that the role of the DSO as an intermediary agent leads to a decrease in operating costs of the VPPs by organizing lower-level trading, while making a profit for itself. However, this result comes with interests loss of the wholesale market, implying that stakeholders in the market need to abide by regulatory constraints.
\end{abstract}


%
\IEEEpeerreviewmaketitle

\section*{Nomenclature}
\addcontentsline{toc}{section}{Nomenclature}

\subsection*{Parameters}
\begin{IEEEdescription}
    \item[$\textrm{T}$] Length of market horizon
     \item[$\textrm{M}$] Constant for `big-M' method (MW)
    \item[$\uplambda^\textrm{CEP}$]  Contractual electricity purchase price (€/MWh)
    \item[$\uplambda^\textrm{CES}$]  Contractual electricity sale price (€/MWh)
    \item[$ \textrm{P}_{\textrm{max}}^{\textrm{VPP,p}}$]  Maximum power purchase volume of VPPs (MWh)\footnote{An hourly clearing of the energy market is assumed for simplicity ($\Delta t=1 \text{h}$), therefore MW and MWh are used interchangeably.}
    \item[$ \textrm{P}_{\textrm{max}}^{\textrm{VPP,s}}$] Maximum power sale volume of VPPs (MWh)
    \vspace{2pt}
    \item[$ \textrm{P}^{\textrm{D}}$]  Fixed demand in VPPs (MW)
    \item[$ \textrm{a},\textrm{b},\textrm{c} $]  Cost coefficients for the micro turbines (€/MWh$^2$, €/MWh, €) 
     \item[$ \textrm{e} $]  Cost coefficient for battery storage (€/MWh$^2$)
     \item[$ \textrm{P}_{\textrm{max}}^{\textrm{MT}} $] Rated power of the micro turbine (MW)
     \vspace{2pt}
     \item[$ \textrm{P}_{\textrm{down}}^{\textrm{MT}}$]  Downwards ramp rate of the micro turbine (MW/h)
     \vspace{2pt}
     \item[$ \textrm{P}_{\textrm{up}}^{\textrm{MT}} $]   Upwards ramp rate of the micro turbine (MW/h)
     \vspace{2pt}
     \item[$ \textrm{P}_{\textrm{max}}^{\textrm{BS}} $]  Rated power of the battery storage device (MW)
     \vspace{2pt}
      \item[$ \textrm{P}_{\textrm{max}}^{\textrm{WT}}$]  Available power from the wind turbines (MW)
      \vspace{2pt}
      \item[$ \textrm{E}_{\textrm{max}}$]  Capacity of the battery storage device (MWh)
       
\end{IEEEdescription}

\subsection*{Decision variables}
\begin{IEEEdescription}
    \item[$\lambda^\textrm{ES}$] Electricity sale price for the VPPs (€/MWh)
    \item[$\lambda^\textrm{EP}$] Electricity purchase price for the VPPs (€/MWh)
     \item[$P^{\textrm{DSO,s}}$]     Electricity sold by DSO to the wholesale power market (MWh)
     \item[$P^{\textrm{DSO,p}}$]     Electricity purchased by DSO from the wholesale power market (MWh)
     \item[$P^{\textrm{VPP,p}}$]     Electricity purchased by VPPs from the DSO (MWh)
     \item[$P^{\textrm{VPP,s}}$]     Electricity sold by VPPs to the DSO (MWh)
      \item[$P^{\textrm{MT}}$]   Output of the micro turbine (MW)
      \item[$P^{\textrm{BS}}$]   Output of the battery storage device (MW)
      \item[$P^{\textrm{WT}}$]   Output of the wind turbines (MW)
      \item[$SoC$]  State of charge of the battery storage device (\%)
      \item[$z$]   Binary variable for the `big-M' method
\end{IEEEdescription}

\section{Introduction} 
Virtual Power Plants (VPPs) are an emerging option for aggregating Distributed Energy Resources (DERs) to the power grid, forming a virtual entity to participate in the electricity trading. In a context where different aggregators participate in the power market, each stakeholder pursues the maximization of its own profit. Some studies have used game theory to analyze the equilibrium problem between stakeholders \cite{WangYao2015,XuZheng2022}.

In previous studies, the role and profit achieved by the Distribution System Operator (DSO) are considered differently \cite{WangYao2015,mu2024distributed,tan2022bi,steriotis2022co,fu2017market}. In \cite{WangYao2015}, the DSO acts as a nonprofitable central controller to manage VPPs in the game. Similarly, the DSO in \cite{mu2024distributed} manages the distribution network operation at the upper level to minimize power losses and enhance the system stability. However, in \cite{tan2022bi}, the DSO maximizes its profits by optimizing the dispatching of the distribution system. The DSO in \cite{steriotis2022co} minimizes its operating costs at the upper level. In \cite{fu2017market}, VPPs can reduce the risks faced by the DSO from uncertainty in the price of the spot market and fluctuations in power consumption, which means that the DSO may benefit indirectly from them.

When the DSO acts as the trading organizer and meanwhile a stakeholder, it should not incur in economic losses, therefore its revenues and costs must be considered in each hourly clearing. In order to study the interactions between the DSO and different VPPs, we assume that the DSO is an intermediary agent for aggregators and sets trading prices for them. The goal is to analyze the behavior of the different selfish agents, to understand the role of the DSO and what insights we can draw from it to have some guidance to the reality.

The DSO and VPPs form a Stackelberg game, also known as a bilevel optimization problem. A bilevel model may be expressed as a Mathematical Program with Equilibrium Constraints (MPEC) under certain conditions \cite{Conejo2020, kim2020mpec}, 
in which the lower-level problems are transformed to be constraints of the upper level. Karush-Kuhn-Tucker (KKT) conditions are the main method of the problem transformation \cite{Yujian2019}, valid if the lower-level problems are convex. This is the method used in the present paper for solving the Stackelberg game.

\section{Stackelberg Game Model for DSO and VPPs}\label{Stackelberg Game Model for DSO and VPPs}
\begin{figure}[t]

    \centering
    \includegraphics[width=0.9\columnwidth]{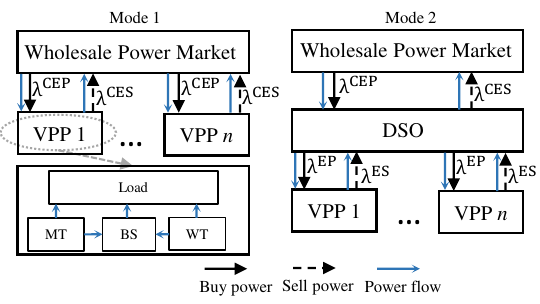}
    \caption{Diagram of energy and cash flows in the two market settings considered.}
    \label{fig:Relationship of energy exchange}
   \vspace{-5mm}
\end{figure}

To minimize operating costs, the internal power balance of a single VPP may not be achieved, showing a surplus or shortage of electricity to the outside grid. 
We consider two different trading possibilities for balancing energy consumption within VPPs, as depicted in Fig.~\ref{fig:Relationship of energy exchange}:
\begin{itemize}
    \item `Mode 1', where VPPs trade directly with the wholesale power market at contract prices $\uplambda^\textrm{CEP}$, $\uplambda^\textrm{CES}$, therefore no game is played. Mathematically, this is a simple setting in which each VPP simply minimizes its own cost, independent of other VPPs.

    \item `Mode 2', where the DSO plays the role of an intermediary agent between the VPPs and the wholesale power market, therefore VPPs sell excess energy to the DSO at price $\lambda^\textrm{ES}$ and buy energy from the DSO at price $\lambda^\textrm{EP}$. Prices are set by the DSO via a Stackelberg game. The DSO also trades with the non-strategic wholesale power market at fixed contract prices $\uplambda^\textrm{CES}$, $\uplambda^\textrm{CEP}$.
\end{itemize}

\begin{figure}[t]
    \centering
    \includegraphics[width=0.7\columnwidth]{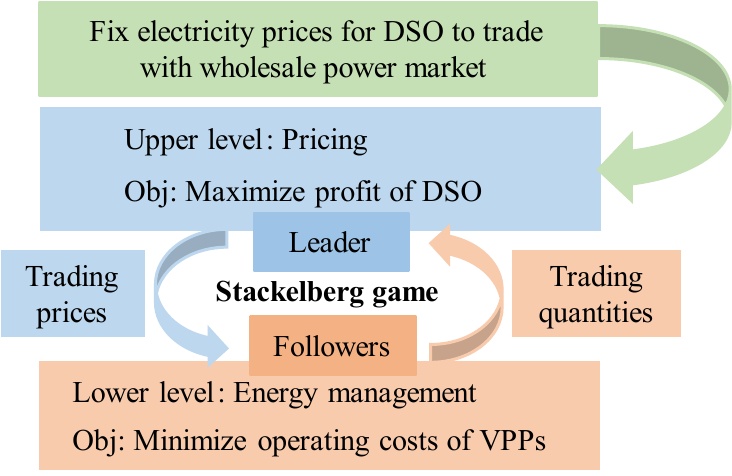}
    \caption{Stackelberg game framework for DSO and VPPs.}
    \label{fig:Stackelberg game framework of DSO and VPPs}
   \vspace{-6mm}
\end{figure}

In Mode 2, the DSO acts as a leader who trades with the wholesale power market and all VPPs, while VPPs are followers who only trade with the DSO. The leader and the followers play a Stackelberg game as described in more detail in Fig. \ref{fig:Stackelberg game framework of DSO and VPPs}. The DSO first agrees with the VPPs on the volume of electricity to be sold or purchased, and then trades with the wholesale power market to realize this traded quantity. The strategic DSO, as the unregulated trading organizer, sets the prices $\lambda^\textrm{ES}$ and $\lambda^\textrm{EP}$ for trading with the VPPs, while guaranteeing that it will not incur a loss when eventually trading with the wholesale market at contract prices $\uplambda^\textrm{CEP}$, $\uplambda^\textrm{CES}$. VPPs take the prices and solve their own economic dispatch, deciding the output of their DERs in order to minimize overall operating costs, therefore eventually setting the volume of electricity traded with the DSO.

Each VPP is considered here to contain dispatchable generation in the form of micro turbines (MTs), weather-driven generation such as wind turbines (WTs), battery storage devices (BS) and a fixed hourly demand representing the energy needs of consumers, as shown in Fig.~\ref{fig:Relationship of energy exchange}.

\subsection{Game Model of Leader-DSO for Pricing} 
The game strategy adopted by DSO is to set $\lambda_t^\textrm{ES}$ and $\lambda_t^\textrm{EP}$ for VPPs, then trade with them. The strategy spaces $\mbox{\boldmath$\lambda$}^{\textrm{ES}}$ and $\mbox{\boldmath$\lambda$}^{\textrm{EP}}$ are, respectively:
\begin{equation}
\label{DSO_strategy_space}
\begin{split}
\mbox{\boldmath$\lambda$}^{\textrm{ES}}=\{\lambda_1^\textrm{ES},\lambda_2^\textrm{ES},...,\lambda_T^\textrm{ES}\}\\
\mbox{\boldmath$\lambda$}^{\textrm{EP}}=\{\lambda_1^\textrm{EP},\lambda_2^\textrm{EP},...,\lambda_T^\textrm{EP}\}
\end{split}
\end{equation}

\subsubsection{Utility Function}
The utility function of DSO is to maximize its profits, which comprises of the cost and revenue from trading with the wholesale market and VPPs:
\begin{equation}\label{eq:DSO_utility}
\begin{split}
\text{max}\quad C^{\textrm{DSO}}=\sum_{t=1}^T (\uplambda_t^\textrm{CES} \cdot P_t^{\textrm{DSO,s}}-\uplambda_t^\textrm{CEP} \cdot P_t^{\textrm{DSO,p}} \\
+\lambda_t^\textrm{EP} \cdot \sum_{j=1}^J P_{j,t}^{\textrm{VPP,p}}-\lambda_t^\textrm{ES} \cdot \sum_{j=1}^J P_{j,t}^{\textrm{VPP,s}})
\end{split}
\end{equation}
Where $\uplambda_t^\textrm{CES} \cdot P_t^{\textrm{DSO,s}}$ and $\lambda_t^\textrm{EP} \cdot \sum_{j=1}^J P_{j,t}^{\textrm{VPP,p}}$ are the income by trading with the wholesale market and VPPs, respectively; $\uplambda_t^\textrm{CEP} \cdot P_t^{\textrm{DSO,p}}$ and $\lambda_t^\textrm{ES} \cdot \sum_{j=1}^J P_{j,t}^{\textrm{VPP,s}}$ are the corresponding transaction costs.
\subsubsection{Constraints}
The premise for aggregators to trade with the DSO is that they can take more favorable prices. Otherwise, for VPPs, this is no different from trading energy directly with the wholesale market. Therefore, the prices set by the DSO for VPPs must meet \eqref{eq:DSO_priceconstraints_1}:
\begin{equation}\label {eq:DSO_priceconstraints_1}
\left \{
\begin{array}{ll}
  \uplambda_t^\textrm{CES} \leq \lambda_t^{\textrm{ES}} \leq \uplambda_t^\textrm{CEP} \\ [4pt]
  \uplambda_t^\textrm{CES} \leq \lambda_t^{\textrm{EP}} \leq \uplambda_t^\textrm{CEP}
\end{array}
\right.
\end{equation}

For transactions with the wholesale market, it is impossible for the DSO to buy and sell electricity at one period, and only one trading decision can be made (buy, sell, or no transaction). We use $P_t^{\textrm{DSO}}$ to indicate the trading for the DSO, based on the demand from each VPP in a unit time. Therefore $P_t^{\textrm{DSO,p}}$ and $P_t^{\textrm{DSO,s}}$ must follow \eqref{eq:DSO_constraints_2}-\eqref{eq:DSO_constraints_4}:
\begin{equation}\label {eq:DSO_constraints_2}
P_t^{\textrm{DSO}}=\sum_{j=1}^J (P_{j,t}^{\textrm{VPP,p}}-P_{j,t}^{\textrm{VPP,s}})
\end{equation}
\begin{equation}\label {eq:DSO_constraints_3}
P_t^{\textrm{DSO,p}} = \left \{
\begin{array}{ll}
  P_t^{\textrm{DSO}}, \quad \textrm{if} \quad P_t^{\textrm{DSO}} \geq 0 \\
  \quad 0,  \quad \quad  \textrm{if}\quad P_t^{\textrm{DSO}} < 0
\end{array}
\right.
\end{equation}
\begin{equation}\label {eq:DSO_constraints_4}
P_t^{\textrm{DSO,s}} = \left \{
\begin{array}{ll}
  -P_t^{\textrm{DSO}}, \; \; \; \textrm{if}\quad P_t^{\textrm{DSO}} < 0 \\
  \quad 0, \quad \quad    \;  \; \textrm{if} \quad P_t^{\textrm{DSO}} \geq 0
\end{array}
\right.
\end{equation}
Where $P_t^{\textrm{DSO}}$ is the quantity of electricity traded by the DSO with the wholesale market, equal to the quantity traded with the VPPs. \eqref{eq:DSO_constraints_2} means that the DSO first clears the market in the lower level by trading with the different VPPs, and then trades with the wholesale market based on the electricity deficit or surplus in the lower level. $P_t^{\textrm{DSO}}\geq0$ indicates that electricity is purchased from the wholesale market by the DSO; while $P_t^{\textrm{DSO}}<0$ means the opposite.

Constraints \eqref{eq:DSO_constraints_3}-\eqref{eq:DSO_constraints_4} involve `if-then' logical statements. A `big-M' method can be used to transform them into indicator constraints containing an auxiliary binary variable, thus making them suitable for optimization solvers. Consider two generic conditional constraints:
\begin{equation}\label {eq:DSO_constraints_5}
\left \{
\begin{array}{ll}
 \quad \mathrm{\textbf{if}}\; \; \; \; \,( x \geq \upalpha), \;  \mathrm{\textbf{then}} \quad y=\text{A} \cdot x \\
 \mathrm{\textbf{else if}} \;\, (x < \upalpha),   \; \mathrm{\textbf{then}} \quad y=\text{B} \cdot x
\end{array}
\right.
\end{equation}
With the binary variable `$z$', \eqref{eq:DSO_constraints_5} can be reformulated as:
\begin{equation}\label {eq:DSO_constraints_6}
\left \{
\begin{array}{ll}
 -\text{M} \cdot (1-z) \leq x-\upalpha \leq \text{M} \cdot z  \\
 -\text{M} \cdot (1-z) \leq y-\text{A} \cdot x \leq \text{M} \cdot (1-z)  \\
 -\text{M} \cdot z \leq y-\text{B} \cdot x \leq \text{M} \cdot z
\end{array}
\right.
\end{equation}
Where if $z=1$, $y=\text{A} \cdot x$; else if $z=0$, $y=\text{B} \cdot x$. Parameter `$\text{M}$' is a constant which should be set as small as possible \cite{williams2013model}, while ensuring that it is also sufficiently high to relax the above constraints appropriately. 

For reformulating \eqref{eq:DSO_constraints_3} in the same way, simply set $\text{A}=1$, $\text{B}=0$, $z_1$, $\upalpha=0$, $x=P_t^{\textrm{DSO}}$, $y=P_t^{\textrm{DSO,p}}$:
\begin{equation}\label {eq:DSO_constraints_7}
\left \{ 
\begin{array}{ll}
 -\text{M} \cdot (1-z_1) \leq P_t^{\textrm{DSO}}-0 \leq \text{M} \cdot z_1 \\[4pt]
 -\text{M} \cdot (1-z_1) \leq P_t^{\textrm{DSO,p}}-P_t^{\textrm{DSO}} \leq \text{M} \cdot (1-z_1)  \\[4pt]
 -\text{M} \cdot z_1 \leq P_t^{\textrm{DSO,p}}-0 \leq \text{M} \cdot z_1
\end{array}
\right.
\end{equation}

Equivalently, using $\text{A}=0$ and $\text{B}=-1$ for \eqref{eq:DSO_constraints_4}:
\begin{equation}\label {eq:DSO_constraints_8}
\left \{
\begin{array}{ll}
 -\text{M} \cdot (1-z_2) \leq P_t^{\textrm{DSO}}-0 \leq \text{M} \cdot z_2 \\[4pt]
 -\text{M} \cdot (1-z_2) \leq P_t^{\textrm{DSO,s}}-0 \leq \text{M} \cdot (1-z_2)  \\[4pt]
 -\text{M} \cdot z_2 \leq P_t^{\textrm{DSO,s}}+P_t^{\textrm{DSO}} \leq \text{M} \cdot z_2
\end{array}
\right.
\end{equation}

In order for \eqref{eq:DSO_constraints_7} and \eqref{eq:DSO_constraints_8} to simultaneously hold at any given period, $z_1$ and $z_2$ must also satisfy:
\begin{equation}
\label {eq:DSO_constraints_9}
z_1=z_2, \quad \forall t
\end{equation}
Note that index `$t$' has been dropped from the binary variables for reasons of clarity in the mathematical expressions.

\subsection{Game Model of Followers-VPPs for Managing Energy}
The game strategy of VPPs is the operating plan, including the volume of electricity traded with the DSO, the output of MT, and the charging and discharging power of BS, all considering the available output of WT and the fixed demand to be satisfied. The strategy space of $\text{VPP}_j$ is then defined as:
\begin{equation}
\label{VPP_strategy_space}
\boldsymbol{\mathit{P}}_j=\{P_{j,t}^{\textrm{VPP,s}},P_{j,t}^{\textrm{VPP,p}},P_{i,t}^{\textrm{MT}},P_{i,t}^{\textrm{BS}},P_{i,t}^{\textrm{WT}}\}
\end{equation}

\subsubsection{Utility Function}
The utility function of VPPs is to minimize operating costs, including the cost of purchasing electricity from DSO, the cost of MT power production given by function `$C_{i,t}^{\textrm{MT}}$', and the degradation cost of BS given by function `$C_{i,t}^{\textrm{BS}}$' due to cycling. Wind power is considered as a completely free resource. This function is given by \eqref{eq:VPP_Utility}:
\begin{equation}\label{eq:VPP_Utility}
\begin{split}
& \text{min}\quad C^{\textrm{VPP}}_j=\sum_{t=1}^T [\lambda_t^\textrm{EP} \cdot P_{j,t}^{\textrm{VPP,p}}-\lambda_t^\textrm{ES} \cdot P_{j,t}^{\textrm{VPP,s}} \\
& \hspace{2cm} +\sum_{i\in \Omega_j}(C_{i,t}^{\textrm{MT}}+C_{i,t}^{\textrm{BS}})]
\end{split}
\end{equation}

\begin{equation}\label{eq:VPP_Utility-MT}
C_{i,t}^{\textrm{MT}}=\textrm{a}_\textit{i} \cdot (P_{i,t}^{\textrm{MT}} \cdot \Delta t)^2+\textrm{b}_\textit{i} \cdot P_{i,t}^{\textrm{MT}}\cdot \Delta t+\textrm{c}_\textit{i}
\end{equation}
\begin{equation}\label{eq:VPP_Utility-ES}
C_{i,t}^{\textrm{BS}}=\textrm{e}_\textit{i} \cdot (P_{i,t}^{\textrm{BS}}\cdot \Delta t)^2
\end{equation}
Where \eqref{eq:VPP_Utility-ES} is composed of the square of charge or discharge, a setting from \cite{lei2016multi}, which makes $C_{i,t}^{\textrm{BS}}$ not change sign in \eqref{eq:VPP_Utility} when $P_{i,t}^{\textrm{BS}}$ is negative.
\subsubsection{Constraints}
When VPPs respond to prices, each of them must meet constraints which are shown in \eqref{eq:VPP_Constraints-1}-\eqref{eq:VPP_Constraints-6}.
\begin{equation}\label{eq:VPP_Constraints-1}
P_{j,t}^{\textrm{VPP}}=P_{j,t}^{\textrm{VPP,p}}-P_{j,t}^{\textrm{VPP,s}}
\end{equation}
\begin{equation}\label {eq:VPP_Constraints-2}
\left \{
\begin{array}{ll}
0 \leq P_{j,t}^{\textrm{VPP,p}} \leq \textrm{P}_{\text{max}}^{\text{VPP,p}}   \\[4pt]
0 \leq P_{j,t}^{\textrm{VPP,s}} \leq \textrm{P}_{\text{max}}^{\text{VPP,s}}
\end{array}
\right.
\end{equation}
\begin{equation}\label{eq:VPP_Constraints-3}
P_{j,t}^{\textrm{VPP}}+\sum_{i\in \Omega_j}(P_{i,t}^{\textrm{MT}}+P_{i,t}^{\textrm{BS}}+P_{i,t}^{\textrm{WT}})\cdot\Delta t=\sum_{i\in \Omega_j}\textrm{P}_{i,t}^{\text{D}} \cdot \Delta t
\end{equation}
\begin{equation}\label {eq:VPP_Constraints-4}
\left \{
\begin{array}{ll}
0 \leq P_{i,t}^{\textrm{MT}} \leq \textrm{P}_{i,\text{max}}^{\text{MT}}   \\[4pt]
\textrm{P}_{i,\text{down}}^{\text{MT}} \cdot \Delta t  \leq P_{i,t+1}^{\textrm{MT}}-P_{i,t}^{\textrm{MT}} \leq \textrm{P}_{i,\text{up}}^{\text{MT}} \cdot \Delta t
\end{array}
\right.
\end{equation}
\begin{equation}\label {eq:VPP_Constraints-5}
\left \{
\begin{array}{ll}
-\textrm{P}_{i,\text{max}}^{\text{BS}} \leq P_{i,t}^{\textrm{BS}} \leq \textrm{P}_{i,\text{max}}^{\text{BS}}   \\[4pt]
SoC_{i,t}= SoC_{i,t-1}-\frac{\Delta t}{\textrm{E}_{i,\text{max}}} \cdot P_{i,t}^{\text{BS}}\\[4pt]
\textrm{SoC}_{i,\text{min}} \leq SoC_{i,t} \leq \textrm{SoC}_{i,\text{max}} \\[4pt]
SoC_{i,0}=SoC_{i,T}
\end{array}
\right.
\end{equation}
\begin{equation}\label{eq:VPP_Constraints-6}
0 \leq P_{i,t}^{\textrm{WT}} \leq \textrm{P}_{i,t,\text{max}}^{\text{WT}}
\end{equation}
\vspace{-0.1cm}
The internal power balance of $\text{VPP}_j$ is represented by \eqref{eq:VPP_Constraints-1}-\eqref{eq:VPP_Constraints-3}. Within a limited transaction volume, VPPs sell or purchase electricity in each unit time to meet the surplus or shortage of electricity. The output and ramp rates of MT are limited by \eqref{eq:VPP_Constraints-4}. The charge and discharge range of the BS and boundary of the SoC are captured by \eqref{eq:VPP_Constraints-5}. We constrain BS to be consistent at the start and end of the cycle to ensure that it has sufficient electricity left to operate in the next cycle. The output of WT is shown in \eqref{eq:VPP_Constraints-6}, indicating that it will not exceed the upper limit of its available resources.

\subsection{Stackelberg Game Model}
Based on the above sections, the Stackelberg game model between the DSO (leader) and multiple aggregators/VPPs (followers) is established in \eqref{eq:bilevel game model}:
\begin{equation}\label{eq:bilevel game model}
\begin{split}
&\mathop{\text{max}}\limits_ {\tiny{\mbox{\boldmath$\lambda$}^{\textrm{ES}},\mbox{\boldmath$\lambda$}^{\textrm{EP}},z_1,z_2}}
C^{\textrm{DSO}}(\mbox{\boldmath$\lambda$}^{\textrm{ES}},\mbox{\boldmath$\lambda$}^{\textrm{EP}},\sum_j\boldsymbol{\mathit{P}}_j,z_1,z_2) \\
& \;\; \; \text{s.t.} \qquad \eqref{eq:DSO_priceconstraints_1}, \eqref{eq:DSO_constraints_2}, \;{\eqref{eq:DSO_constraints_7}-\eqref{eq:DSO_constraints_9}}\\[4pt]
& \hspace{12mm} \left \{
\begin{array}{ll}
\mathop{\text{min}}\limits_{\tiny{\boldsymbol{\mathit{P}}_j}} \quad \sum\limits_j C_j^{\textrm{VPP}}(\mbox{\boldmath$\lambda$}^\textrm{ES},\mbox{\boldmath$\lambda$}^{\textrm{EP}},\boldsymbol{\mathit{P}}_j)\\
\text{s.t.} \quad{\eqref{eq:VPP_Constraints-1}-\eqref{eq:VPP_Constraints-6}}
\end{array}
\right\}
\end{split}
\end{equation}
Where DSO and VPPs formulate strategies with the goal of maximizing profits and minimizing operating costs, respectively. The DSO profit is related to $\mbox{\boldmath$\lambda$}^{\textrm{ES}}$, $\mbox{\boldmath$\lambda$}^{\textrm{EP}}$ and the transaction quantities $P_{j,t}^{\textrm{VPP,s}}$, $P_{j,t}^{\textrm{VPP,p}}$ in $\boldsymbol{\mathit{P}}_j,\forall j$ (as defined in \eqref{VPP_strategy_space}). The responses of VPPs to electricity prices affect the profit of DSO: high $\mbox{\boldmath$\lambda$}^{\textrm{EP}}$ results in a low purchased quantity by VPPs, while low $\mbox{\boldmath$\lambda$}^{\textrm{ES}}$, implies a low volume of electricity sold by VPPs. Therefore, the DSO has to consider these movements and find the Nash equilibrium solution as the optimal prices. 
The definition and proof of the equilibrium solution for the Stackelberg game are explained in \cite{Conejo2020}.
  \begin{table}[t]
   \captionsetup{ textfont={sc,footnotesize}, labelfont=footnotesize, labelsep=newline} 
  \renewcommand{\arraystretch}{1.5}
  \caption{Parameters Setting of Each VPP}
  \centering
  \resizebox{\columnwidth}{!}{
  \begin{tabular}{c|c|c|c}
  \multicolumn{1}{c|}{Parameters} & $ \text{VPP}_\text{1}$ & $ \text{VPP}_\text{2}$ & $ \text{VPP}_\text{3}$ \\  \hline 
  $\textrm{P}_{\textrm{max}}^{\textrm{VPP,p}},  \textrm{P}_{\textrm{max}}^{\textrm{VPP,s}}$                       & 10, 10    & 10, 10      & 10, 10   \\ 
  $ \textrm{a},\textrm{b},\textrm{c} $                      & 0.08, 0.90, 1.20       & 0.1, 0.6, 1.0         &  0.15, 0.50, 0.80   \\ 
  $ \textrm{e} $                                       & 0.05      & 0.05       & 0.05    \\ 
  $  \textrm{P}_{\text{max}}^{\text{MT}},\textrm{P}_{\text{down}}^{\text{MT}},\textrm{P}_{\text{up}}^{\text{MT}}$                      &6.0, -3.5, 3.5   & 5, -3, 3     &4, -2, 2   \\ 
  $ \textrm{P}_{\text{max}}^{\text{BS}}$                   & 0.6       & 0.6          & 1.2   \\ 
  $\textrm{SoC}_{\textrm{0}},\textrm{SoC}_{\text{min}},\textrm{SoC}_{\text{max}}$                     &40, 20, 90      &40, 20, 90     &40, 20, 90    \\ 
  $\textrm{E}_{\text{max}}$                  & 1    & 1      & 2  \\ \hline
  \end{tabular}%
  \label{tab:System parameters setting}
  }
  \end{table}
\begin{figure}[!t]
    \centering
    \begin{subfigure}
        \centering
        \includegraphics[width=122pt]{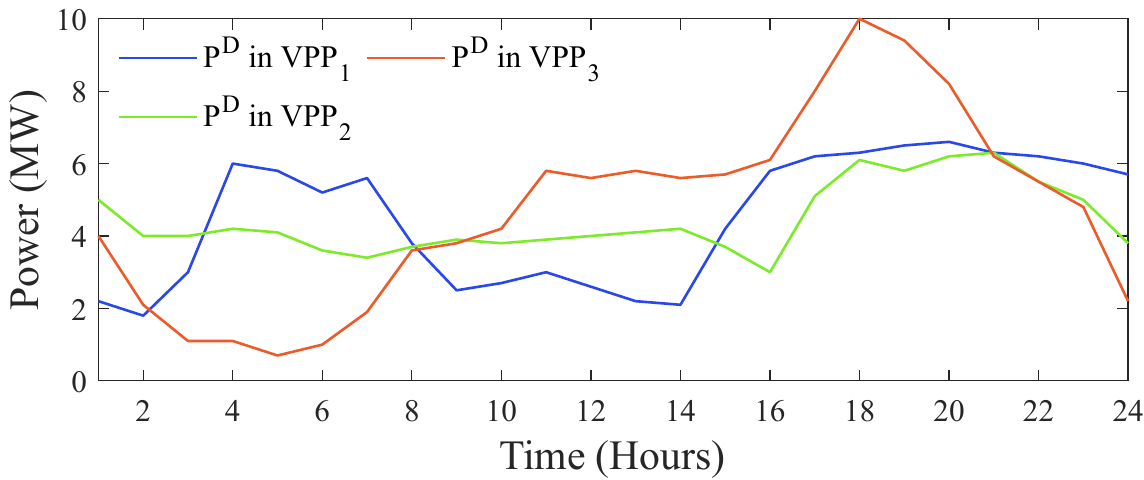}
    \end{subfigure}
    \hfill
    \begin{subfigure}
        \centering
       \includegraphics[width=122pt]{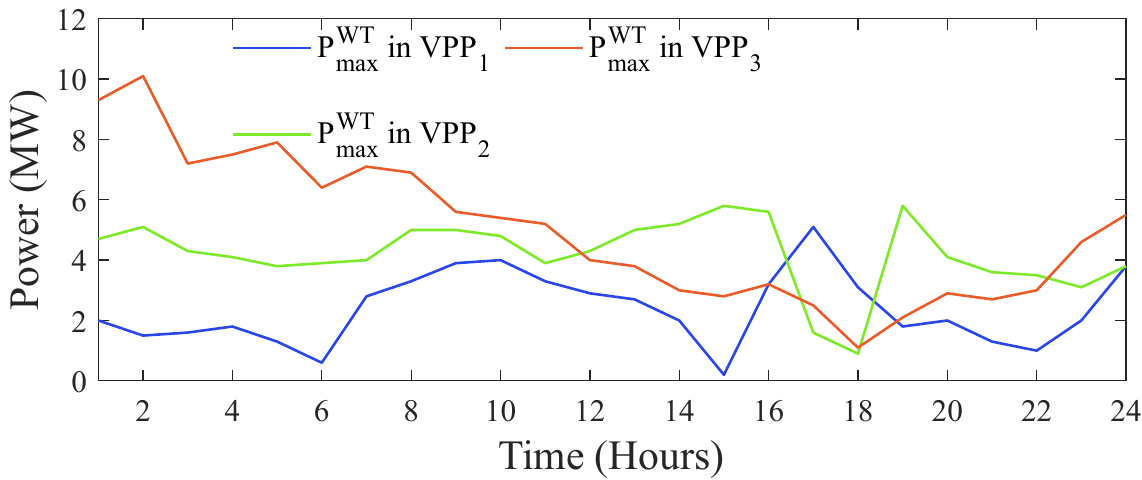}
    \end{subfigure}
    \vspace{-3mm}
    \caption{Load profiles and available wind power in each VPP.}
    \label{fig:wt_and_load_in_each_VPP}
    \vspace{-3.5mm}
\end{figure}
\section{Case Studies}\label{Case Studies}
 The above game model in \eqref{eq:bilevel game model} is a bilevel problem, where the upper level is a mixed-integer nonlinear program. Its non-linearity comes from the products of decision variables $\lambda_t^\textrm{EP} \cdot \sum_{j=1}^J P_{j,t}^{\textrm{VPP,p}}$ and $\lambda_t^\textrm{ES} \cdot \sum_{j=1}^J P_{j,t}^{\textrm{VPP,s}}$ in \eqref{eq:DSO_utility}. The lower level, i.e., \eqref{eq:VPP_Utility}-\eqref{eq:VPP_Constraints-6}, is a convex nonlinear program. Therefore, the lower level can be exactly replaced by its corresponding KKT optimality conditions \cite{Conejo2020}, transforming the original bilevel problem into a single-level MPEC:
\begin{equation}\label{eq:single-level model}\
\begin{split}
&\text{max}\quad C^{\textrm{DSO}}(\cdot)\\
&\text{s.t.}  \quad\left \{ \eqref{eq:DSO_priceconstraints_1},\eqref{eq:DSO_constraints_2},\;\eqref{eq:DSO_constraints_7}-\eqref{eq:DSO_constraints_9},\;\textrm{KKT}^{(j)} \;\forall j\right\} 
\end{split}
\end{equation}

Detailed derivations and explanations on MPEC and KKT conditions can be found in \cite{Conejo2020}. Here, \texttt{BilevelJuMP.jl} is used to compile the KKT conditions, in order to avoid the need to write them down manually. This is a Julia package that enables the user to describe both upper- and lower-level problems using the \texttt{JuMP.jl} algebraic syntax \cite{Garcia2024}. In the simulation, we choose \texttt{SCIP.jl} as the solver due to its compatibility with mixed-integer nonlinear programs \cite{BestuzhevaChmielaMuellerSerranoVigerskeWegscheider2023}, which leads to feasible solutions. The code used to run the simulations is available on a public repository \cite{Code}. 

\subsection{Test System Setting}
The test system comprises three VPPs of similar size. The primary differences among the VPPs lie in their distinct operating parameters of the DER as shown in Table \ref{tab:System parameters setting}, and the internal demand profiles and available wind power, as shown in Fig. \ref{fig:wt_and_load_in_each_VPP}. For MT, each VPP has unique cost coefficients (a, b, c), output upper limits ($ \textrm{P}_{\textrm{max}}^{\textrm{MT}}$), and ramp rates ($\textrm{P}_{\textrm{up}}^{\textrm{MT}}$, $\textrm{P}_{\textrm{down}}^{\textrm{MT}}$). In terms of BS, the variations include the capacity ($ \textrm{E}_{\textrm{max}}$) and the upper limits for charge and discharge ($ \textrm{P}_{\textrm{max}}^{\textrm{BS}} $). Furthermore, the power available from the WT ($ \textrm{P}_{\textrm{max}}^{\textrm{WT}}$) differs between VPPs. Therefore, each VPP may have different dispatch strategies based on these various system settings.

The maximum transaction volume ($ \textrm{P}_{\textrm{max}}^{\textrm{VPP,p}}$, $ \textrm{P}_{\textrm{max}}^{\textrm{VPP,s}}$) of each single VPP is 10 MWh, i.e., the upper bound of the trading volumes for three VPPs is 30 MWh. Therefore, a tight value for the `big-M' method is provided in \eqref{eq:DSO_constraints_7}-\eqref{eq:DSO_constraints_9} of $\textrm{M}=30$.

\subsection{Analysis of Trading Results through the Stackelberg Game}
\begin{table}[t]
    \captionsetup{justification=centering, textfont={sc,footnotesize}, labelfont=footnotesize, labelsep=newline} 
    \renewcommand{\arraystretch}{1.2}
    \centering
    \caption{ Costs and Profits of Stakeholders in Two Modes} 
    \label{tab:Interests of VPPs and DSO}
    \setlength{\tabcolsep}{0mm} 
    \begin{tabular}{
        >{\centering\arraybackslash}p{2.9cm}|
        >{\centering\arraybackslash}p{2.9cm}|
        >{\centering\arraybackslash}p{2.9cm}
    }
     Stakeholders    &Mode 1 (k\text{€}) &Mode 2 (k\text{€}) \\ \hline
    $\text{VPP}_\text{1}  $      &$-3.947$   & $-3.942$           \\
    $\text{VPP}_\text{2} $       & $-0.918$  & $-0.891$            \\
    $\text{VPP}_\text{3}  $       & $-3.587$   & $-3.559$         \\
    $ \text{DSO}  $     & ---   & $1.134$       \\
    $ \text{Wholesale market} $     & $5.370$ & $4.152$     \\
    \hline
    \end{tabular}
\end{table}

\begin{figure}[t]
    \centering
    \includegraphics[width=1\columnwidth]{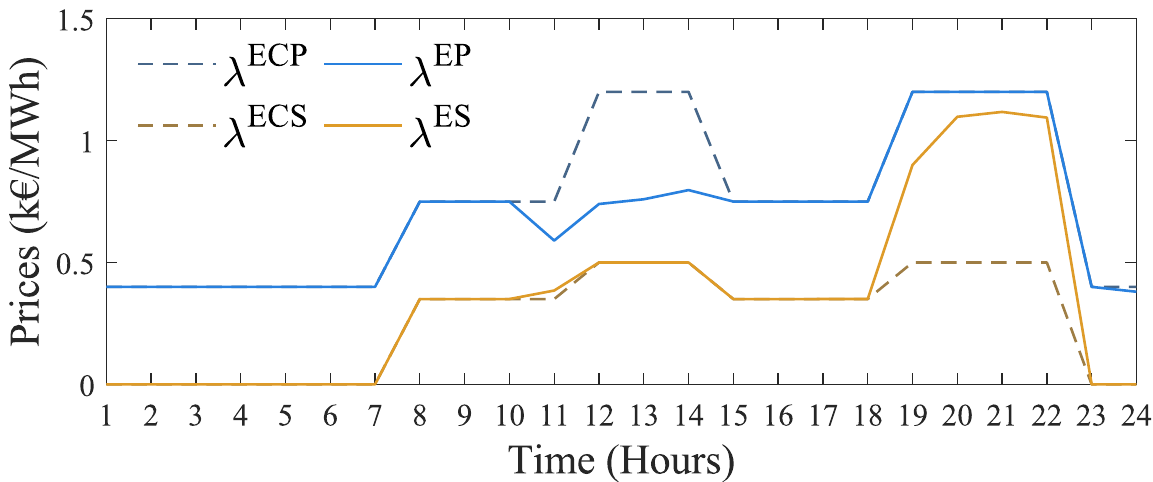}
    \caption{{Trading prices in two modes.}}
    \label{fig:prices}
    \vspace{-5mm}
\end{figure}

Here we examine the strategic behaviors of the DSO and VPPs under the two market modes in Section~\ref{Stackelberg Game Model for DSO and VPPs}. The profits of the DSO, and costs of each VPP, are shown in Table \ref{tab:Interests of VPPs and DSO}.

In Mode 2, compared with Mode 1, the DSO earns profits by setting hourly prices $\lambda_t^{\textrm{ES}}$, $\lambda_t^{\textrm{EP}}$, as shown in Fig. \ref{fig:prices}, and trading with VPPs. The costs of all VPPs decrease ($0.13\%$, $2.94\%$, and $0.78\%$, respectively). The reason is that the revenues flowing into the wholesale market are $22.68\%$ lower, which is almost equal to the sum of the decreased costs of the VPPs and the profits of the DSO. This result shows that selfish DSO and VPPs strategically expand their benefits in the game, by decreasing the net economic flows to the non-strategic wholesale power market.

To clarify how the inflow of money to the wholesale power market decreases, it is necessary to analyze the differences in the solutions for the two market modes, i.e., $\{\boldsymbol{\mathit{P}}_j,\mbox{\boldmath$\lambda$}^{\textrm{ES}}, \mbox{\boldmath$\lambda$}^{\textrm{EP}}\}_{\text{Mode 2}}-\{\boldsymbol{\mathit{P}}_j,\uplambda^\textrm{CES}, \uplambda^\textrm{CEP}\}_{\text{Mode 1}},{\forall j}$ (shown as `Trade diff' and `output diff' in Tables~\ref{tab:Comparison of solutions under different strategies for VPP_1} to \ref{tab:Comparison of solutions under different strategies for VPP_3}). We individualize the analysis for each VPP in the following subsections.

\subsubsection{Cost Reduction Analysis for $\text{VPP}_\text{1}$}
The operations of WT and MT are the same in the two market modes, which means the cost reduction for $\text{VPP}_\text{1}$ comes from the difference in volume of electricity traded, operation of BS, and prices. Table \ref{tab:Comparison of solutions under different strategies for VPP_1} only lists the hours which show a different solution between modes 1 and 
 2, omitting hours with no change at all.
 
\begin{table}[t]
 \captionsetup{justification=centering, textfont={sc,footnotesize}, labelfont=footnotesize, labelsep=newline} 
 {\fontsize{6pt}{9pt}\selectfont
    \centering
    \caption{Comparison of Solutions under Different Modes for $ \text{VPP}_\text{1}$}
    \label{tab:Comparison of solutions under different strategies for VPP_1}
\begin{tabular}{cc|c|c|c|c|c|c}
\multicolumn{2}{c|}{Time (Hours)}                                                                             & 9     & 10    & 11   & $\cdots$ & 23    & 24   \\ \hline
\multicolumn{1}{c|}{\multirow{2}{*}{\begin{tabular}[c]{@{}c@{}}Trade diff\\ (MWh)\end{tabular}}} & Purchase   & ---   & ---   & ---  & $\cdots$ & $\mathbf{-0.10}$ & $\mathbf{0.10}$ \\ \cline{2-2}
\multicolumn{1}{c|}{}                                                                           & Sale       & $-0.12$ & $-0.12$ & $\mathbf{-0.23}$ & $\cdots$ & ---   & ---  \\ \hline
\multicolumn{1}{c|}{\multirow{2}{*}{SoC (\%)}}                                                       & Mode 2 & $55$  & $90$  & $90$  & $\cdots$ & $20$  & $40$  \\ \cline{2-2}
\multicolumn{1}{c|}{}                                                                           & Mode 1 & $43$  & $67$  & $90$  & $\cdots$ & $30$  & $40$  \\ \hline
\end{tabular}
}
\end{table}

Combined with Fig. \ref{fig:prices}, it can be found that $\text{VPP}_\text{1}$ responds to trading prices to reduce operating costs. The sale price $\lambda_{11}^\textrm{ES}$ (where `11' means $t=$11:00), set by DSO, is higher than $\uplambda_{11}^\textrm{CES}$, so more electricity is sold. But in Mode 1, without DSO, $\text{VPP}_\text{1}$ chooses to sell more electricity at 09:00 and 10:00 when the sale price is cheaper. The purchase price $\lambda_{23}^\textrm{EP}$ is higher than $\lambda_{24}^\textrm{EP}$, so less electricity is bought at 23:00 in Mode 2. $\text{VPP}_\text{1}$ strategically reduces costs by charging and discharging the BS: purchasing electricity at a lower price, the SoC increases; selling electricity at a higher price, SoC decreases.

\subsubsection{Cost Reduction Analysis for $\text{VPP}_\text{2}$}
All DER show operating differences under the two modes. Table \ref{tab:Comparison of solutions under different strategies for VPP_2} in the Appendix lists the different solutions between the two modes. The purchase prices $\lambda_{11}^\textrm{EP}$-$\lambda_{14}^\textrm{EP}$, set by the DSO, are lower than the contract price $\uplambda^\textrm{CEP}$, and the purchase quantities by $\text{VPP}_\text{2}$ under the two modes are the same, the costs are thereby reduced in Mode 2. At 17:00 and 18:00, $\text{VPP}_\text{2}$ in Mode 2 chooses to purchase more electricity to sell, at an expensive price, from 19:00 to 22:00, to increase revenue. The price $\lambda_{24}^\textrm{EP}$ is lower than $\lambda_{23}^\textrm{EP}$, so $\text{VPP}_\text{2}$ chooses to purchase more energy.

\begin{figure}[t]
    \centering
    \includegraphics[width=1\columnwidth]{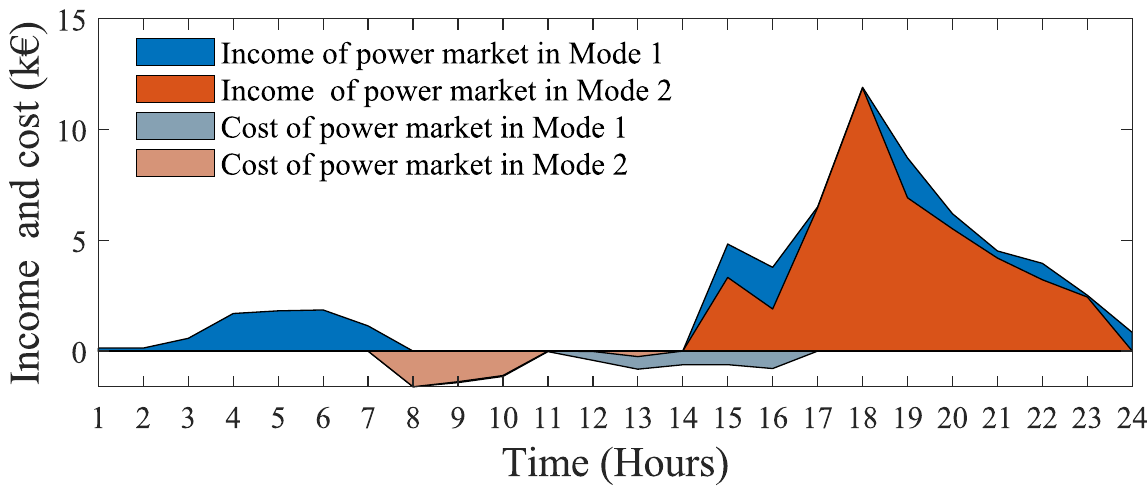}
    \caption{Profits changing of the wholesale power market.}
    \label{fig:pmprofitandcist}
    \vspace{-6mm}
\end{figure}
The costs reduction for $\text{VPP}_\text{2}$ requires the strategic operation of the internal DER. From 02:00 to 07:00, the operations of MT and BS under the two modes are the same, the output of WT directly affects the transactions. From 19:00 to 22:00, $\text{VPP}_\text{2}$ trades with the DSO by discharging the BS and increasing the MT output, thus increasing income. At 23:00, under Mode 1, more electricity is purchased to increase SoC, but this is not as cost-effective as trading at 24:00 in Mode 2. 

\subsubsection{Cost Reduction Analysis for $\text{VPP}_\text{3}$}
Table \ref{tab:Comparison of solutions under different strategies for VPP_3} in the Appendix lists the different solutions between the two modes. $\text{VPP}_\text{3}$ strategically purchases more electricity from the DSO when prices $\lambda_{11}^\textrm{EP}$-$\lambda_{14}^\textrm{EP}$ are cheaper than contract prices, and reduces the purchased volume in the following four hours, when purchase prices rise. From 11:00-14:00, MT reduces output, and internal power balance is maintained mainly by the purchased energy.

\subsubsection{Profit Analysis for Wholesale Market and DSO}
\begin{figure}[t]
    \centering
    \includegraphics[width=1\columnwidth]{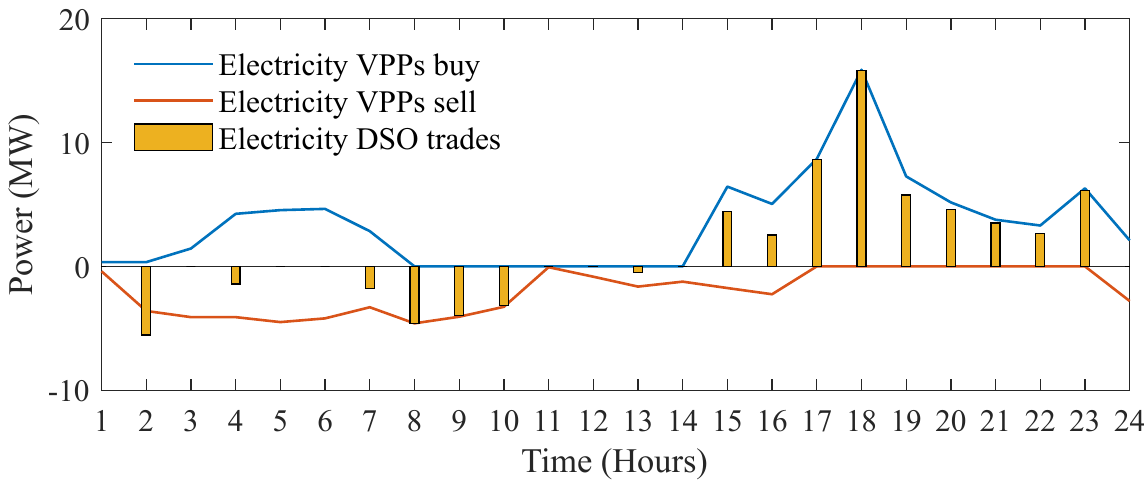}
    \caption{Quantities DSO and VPPs trade with the wholesale power market respectively.}
    \label{fig:vpps_dso_trade}
    \vspace{-5mm}
\end{figure}
The revenue and cost for the wholesale market are shown in Fig. \ref{fig:pmprofitandcist}. 
The income in Mode 2 (selling electricity to DSO) is significantly less than that in Mode 1 (selling electricity directly to VPPs). This leads to a reduction in money flowing into the wholesale market, where the loss is caused by the trading quantity (as the contract prices are fixed, not variables). This same volume of electricity is instead traded in the lower level between the VPPs. 

The volumes that the DSO and VPPs trade with the wholesale market respectively in Mode 2 and Mode 1 are shown in Fig. \ref{fig:vpps_dso_trade}. The negative half-axis represents the sale of electricity to the wholesale market, and the positive half-axis represents the purchase of electricity from the wholesale market.

As the intermediary agent for all aggregators, the DSO trades with the wholesale market at contract prices ($\uplambda^\textrm{CEP}$, $\uplambda^\textrm{CES}$) on the one hand, and sets more favorable `internal prices' ($\lambda^\textrm{EP}$, $\lambda^\textrm{ES}$) based on the operation of each VPP, on the other hand. The DSO directly trades with the VPPs, that is, the lower level forms a trading cluster, resulting in a decrease in volumes traded with the wholesale market. Compared with Mode 1, the DSO promotes power sharing in (\ref{eq:DSO_constraints_2}) between VPPs at the `internal prices', making a profit in the process. This profit results in lower revenue flowing into the wholesale market, as can be seen in Fig. \ref{fig:vpps_dso_trade}: Before 08:00 and after 15:00, the results are significantly different for the cases in which VPPs trade directly with the wholesale market vs.~when the DSO is in between. The selfish aggregators take the prices set by the DSO and prioritize trading with other aggregators, resulting in energy balancing happening mostly at DSO level. 

As shown in Table \ref{tab:Interests of VPPs and DSO} and the previous analysis, each stakeholder is pursuing the favorable profits and costs, leading to the reduced economic flowing into the wholesale market, which are equal to the profits gained by the DSO, plus the cost reduction for the VPPs. Absolutely, this is unfair to the wholesale market due to the absence of proper market regulation. Therefore, in the real-world market, it is significant to design a market regulation and make stakeholders comply to maintain market fairness. Only in this way will market participants have the motivation to provide services. 

\newpage
\section{Conclusion}\label{Conclusion}
This work provides a detailed analysis of the interactions between the DSO and VPPs in electricity trading through a Stackelberg game. The findings highlight that adopting a bi-level model allows the DSO, acting as the intermediary agent, to set trading prices that reduce operating costs for VPPs, while ensuring profitability for itself. From this we also learn that proper regulation is needed to secure a fair market.

While these findings are based on results for a particular setting, i.e., a small test case with three VPPs, the promising trends identified justify carrying out additional theoretical work in the future. The advantages of DSO-VPP trading in a general market setting should be further investigated via mathematical proofs, in order to enhance interest in market participation of these resource aggregators. 

In future work, we will investigate market settings for ancillary services, exploring the coupling relationship between the electricity market and the stable operation of the system.
\vspace{0.5cm}
\section{Acknowledgement}
This work was supported by MICIU/AEI/10.13039/501100011033 and ERDF/EU under grant PID2023-150401OA-C22, as well as by the Madrid Government (Comunidad de Madrid-Spain) under the Multiannual Agreement 2023-2026 with Universidad Politécnica de Madrid, ``Line A - Emerging PIs''. The work of Peng Wang was also supported by China Scholarship Council under grant 202408500065.
\vspace{0.5cm}
\appendix 
Tables \ref{tab:Comparison of solutions under different strategies for VPP_2} and \ref{tab:Comparison of solutions under different strategies for VPP_3} list the different solutions between two modes, for $\textrm{VPP}_\textrm{2}$ and $\textrm{VPP}_\textrm{3}$, respectively.

\begin{table}[h]
 \captionsetup{justification=centering, textfont={sc,footnotesize}, labelfont=footnotesize, labelsep=newline} 
 \caption{Comparison of Solutions under Different Modes for $ \text{VPP}_\text{2}$}
 {\fontsize{6pt}{9pt}\selectfont
    \centering
    \label{tab:Comparison of solutions under different strategies for VPP_2}
     \resizebox{\textwidth}{!}{
\begin{tabular}{cc|c|c|c|c|c|c|c|c|c|c|c|c|c|c}
\multicolumn{2}{c|}{Time (Hours)}                                  & 2     & 3    & $\cdots$ & 6    & 7    & $\cdots$ & 17   & 18   & 19   & 20   & 21   & 22   & 23    & 24   \\ \hline
\multicolumn{1}{c|}{\multirow{2}{*}{Trade diff (MWh)}} & Purchase   & ---   & ---  & $\cdots$ & ---  & ---  & $\cdots$ & $\mathbf{0.21}$ & $\mathbf{0.21}$ & ---  & ---  & ---  & ---  & $\mathbf{-0.10}$ & $\mathbf{0.10}$ \\ \cline{2-2}
\multicolumn{1}{c|}{}                                & Sale       & $-0.04$ & $0.01$ & $\cdots$ & $0.01$ & $0.19$ & $\cdots$ & ---  & ---  & $\mathbf{1.50}$ & $\mathbf{0.57}$ & $\mathbf{0.27}$ & $\mathbf{0.62}$ & ---   & ---  \\ \hline
\multicolumn{2}{c|}{MT output diff (MW)}                           & $0$   & $0$  & $\cdots$ & $0$  & $0$  & $\cdots$ & $0$  & $0$  & $1.07$ & $0.51$ & $0.41$ & $0.53$ & $0$   & $0$  \\ \hline
\multicolumn{2}{c|}{WT output diff (MW)}                           & $\mathbf{-0.04}$ & $\mathbf{0.01}$ & $\cdots$ & $\mathbf{0.01}$ & $\mathbf{0.19}$ & $\cdots$ & $0$  & $0$  & $0$  & $0$  & $0$  & $0$  & $0$   & $0$  \\ \hline
\multicolumn{1}{c|}{\multirow{2}{*}{SoC (\%)}}            & Mode 2 & $48$   & $75$  & $\cdots$ & $63$  & $90$  & $\cdots$ & $90$ & $90$ & $90$  & $72$ & $34$ & $20$  & $20$  & $40$  \\ \cline{2-2}
\multicolumn{1}{c|}{}                                & Mode 1 & $48$   & $75$  & $\cdots$ & $63$  & $90$  & $\cdots$ & $69$ & $47$ & $90$  & $78$ & $26$ & $20$  & $30$  & $40$  \\ \hline
\end{tabular}
}
}

\end{table}

\begin{table}[h]
\captionsetup{justification=centering, textfont={sc,footnotesize}, labelfont=footnotesize, labelsep=newline} 
 {\fontsize{6pt}{9pt}\selectfont
    \centering
    \caption{Comparison of Solutions under Different Modes for $ \text{VPP}_\text{3}$}
    \label{tab:Comparison of solutions under different strategies for VPP_3}
    \resizebox{\textwidth}{!}{
\begin{tabular}{cc|c|c|c|c|c|c|c|c|c|c|c|c|c|c|c|c|c|c}
\multicolumn{2}{c|}{Time (Hours)}                                  & 1     & 2    & 3     & 4    & 5    & 6    & 7    & $\cdots$ & 11    & 12    & 13    & 14    & 15    & 16    & 17    & 18    & $\cdots$ & 24    \\ \hline
\multicolumn{1}{c|}{\multirow{2}{*}{Trade diff (MWh)}} & Purchase   & ---     & ---    & ---     & ---    & ---    & ---    & ---    & $\cdots$ & $\mathbf{0.30}$  & $\mathbf{0.80}$  & $\mathbf{1.14}$  & $\mathbf{1.23}$  & $\mathbf{-0.26}$ & $\mathbf{-0.26}$ & $\mathbf{-0.26}$ & $\mathbf{-0.26}$ & $\cdots$ & ---     \\ \cline{2-2}
\multicolumn{1}{c|}{}                                & Sale       & $-0.06$ & $2.32$ & $-2.67$ & $1.56$ & $0.04$ & $0.43$ & $1.16$ & $\cdots$ & ---     & ---     & ---     & ---     & ---     & ---     & ---     & ---     & $\cdots$ & $-0.50$ \\ \hline
\multicolumn{2}{c|}{MT output diff (MW)}                           & $0$     & $0$    & $0$     & $0$    & $0$    & $0$    & $0$    & $\cdots$ & $\mathbf{-0.30}$ & $\mathbf{-0.68}$ & $\mathbf{-0.72}$ & $\mathbf{-0.74}$ & $0$     & $0$     & $0$  & $0$     & $\cdots$ & $0$     \\ \hline
\multicolumn{2}{c|}{WT output diff (MW)}                           & $-0.06$ & $2.32$ & $-2.67$ & $1.56$ & $0.04$ & $0.43$ & $1.16$ & $\cdots$ & $0$     & $0$     & $0$     & $0$     & $0$     & $0$     & $0$     & $0$     & $\cdots$ & $-0.50$ \\ \hline
\multicolumn{1}{c|}{\multirow{2}{*}{SoC (\%)}}            & Mode 2 & $47$  & $54$ & $61$  & $69$ & $76$ & $83$ & $90$ & $\cdots$ & $90$  & $90$  & $90$  & $71$  & $76$  & $81$  & $85$  & $90$  & $\cdots$ & $40$  \\ \cline{2-2}
\multicolumn{1}{c|}{}                                & Mode 1 & $47$  & $54$ & $61$  & $69$ & $76$ & $83$ & $90$ & $\cdots$ & $90$  & $84$  & $63$  & $20$  & $37$  & $55$  & $73$  & $90$  & $\cdots$ & $40$  \\ \hline
\end{tabular}
}
}
\end{table}

\newpage
\bibliographystyle{IEEEtran} 
\bibliography{main}

\end{document}